\documentclass{article}

\usepackage{arxiv}

\usepackage[utf8]{inputenc} 

\usepackage[T1]{fontenc}    
\usepackage{hyperref}       
\usepackage{url}            
\usepackage{booktabs}       
\usepackage{amsfonts}       
\usepackage{nicefrac}       
\usepackage{microtype}      
\usepackage{lipsum}
\usepackage{graphicx}
\usepackage{multirow}
\usepackage{amsmath}
\usepackage[table]{xcolor}

\graphicspath{ {./images/} }

\title{Comparative Analysis of Audio Feature Extraction for Real-Time Talking Portrait Synthesis}

\author{
 Pegah Salehi  \\
  SimulaMet \\
  Oslo, Norway\\
  \texttt{pegah@simula.no} \\
   \And
 Sajad Amouei Sheshkal  \\
  SimulaMet \\
  Oslo, Norway\\
  \texttt{sajad.amouei@gmail.com} \\
  \And
 Vajira Thambawita  \\
  SimulaMet \\
  Oslo, Norway\\
  \texttt{vajira@simula.no} \\
  \And
 Sushant Gautam \\
  SimulaMet \\
  Oslo, Norway\\
  \texttt{sushant@simula.no} \\
  \And
 Saeed S. Sabet \\
  Forzasys \\
  Oslo, Norway\\
  \texttt{saeed.sh.sabet@gmail.com} \\
  \And
 Dag Johansen \\
  The University of Tromsø \\
  Tromsø, Norway\\
  \texttt{dag.johansen@uit.no} \\
  \And
 Michael A. Riegler \\
  SimulaMet \\
  Oslo, Norway\\
  \texttt{michael@simula.no} \\
  \And
 Pål Halvorsen \\
  SimulaMet \\
  Oslo, Norway\\
  \texttt{paalh@simula.no} \\
}

\begin{document}
\maketitle
\begin{abstract}
This paper examines the integration of real-time talking-head generation for interviewer training, focusing on overcoming challenges in Audio Feature Extraction (AFE), which often introduces latency and limits responsiveness in real-time applications. To address these issues, we propose and implement a fully integrated system that replaces conventional AFE models with Open AI's Whisper, leveraging its encoder to optimize processing and improve overall system efficiency. Our evaluation of two open-source real-time models across three different datasets shows that Whisper not only accelerates processing but also improves specific aspects of rendering quality, resulting in more realistic and responsive talking-head interactions. These advancements make the system a more effective tool for immersive, interactive training applications, expanding the potential of AI-driven avatars in interviewer training.
\end{abstract}



\section{Introduction}

The application of AI in education has gained widespread attention for its potential to enhance learning experiences across disciplines, including psychology~\cite{chernikova2020simulation, crompton2020psychological}. 
In the context of investigative interviewing, especially when questioning suspected child victims, AI offers a promising alternative to traditional training approaches. These conventional methods, often delivered through short workshops, fail to provide the hands-on practice, feedback, and continuous engagement needed for interviewers to master best practices in questioning child victims~\cite{lamb2016difficulties, powell2008designing}. Research has shown that while best practices recommend open-ended questions and discourage leading or suggestive queries~\cite{lamb2007structured, lyon2014interviewing}, many interviewers still struggle to implement these techniques effectively during real-world investigations~\cite{lamb2018tell}. The adoption of AI-powered child avatars provides a valuable solution, enabling Child Protective Services (CPS) workers to engage in realistic practice sessions without the ethical dilemmas associated with using real children, while simultaneously offering personalized feedback on their performance~\cite{powell2022overview}.

Our current system leverages advanced AI techniques within a structured virtual environment to train professionals in investigative interviewing. Specifically, this system integrates the Unity Engine to generate virtual avatars. Despite the potential advantages of our AI-based training system, its effectiveness largely depends on the perceived realism and fidelity of the virtual avatars used in these simulations~\cite{hassan2022towards}. Based on our findings, we observed that avatars generated using Generative Adversarial Networks (GANs) demonstrated higher levels of realism compared to those created with the Unity Engine in several key aspects~\cite{salehi2022more}. 

Accordingly, in this paper, we propose leveraging existing real-time talking portrait generation techniques to create an interactive avatar and evaluate its potential for improving perceived realism and interaction quality. 

Audio-driven talking portrait synthesis enables the animation of a specific person based on arbitrary speech input, using deep learning-based methods such as Neural Radiance Fields (NeRF)~\cite{mildenhall2021nerf} and 3D Morphable Models (3DMMs) to generate high-quality 3D head models that offer flexibility in head poses and superior visual fidelity. Despite advancements in the development of real-time talking head systems, their broad availability remains constrained by the inherent complexity of integrating multi-modal data inputs, including audio and visual cues such as facial landmarks. This complexity poses challenges in achieving both synchronization and processing efficiency, which are imperative for the smooth operation of these systems in real-time scenarios.

In audio-driven talking-head synthesis, the nuances of audio features directly influence the realism and synchronization of the generated visuals. Since different AFE models capture slightly varied aspects from the same audio stream, selecting the most effective model becomes essential. Through our exploration of open-source real-time talking head models, we observed that latency in AFE can hinder real-time performance, impacting the overall responsiveness and realism of the avatar. To mitigate the latency issues associated with existing Audio Feature Extraction (AFE) models, we have leveraged Whisper~\cite{radford2023robust}, an advanced automatic speech recognition (ASR) system that has been adapted for AFE tasks. Trained on extensive multilingual and multitask datasets, Whisper offers an efficient and accelerated solution for AFE, potentially optimizing the real-time capabilities of talking head systems.

In summary, our main contributions are as follows:
\begin{itemize}
    \item Integrating a complete AI-based avatar system combining GPT-3 for conversations, Speech-To-Text (STT), Text-To-Speech (TTS), AFE, and talking portrait synthesis (Figure~\ref{fig:arch}) for full end-to-end experiments.
    \item Evaluating and comparing talking-head synthesis models.
    \item Evaluation and comparison of four AFE models using two open-source talking-head frameworks across three datasets.
    \item Modifying Whisper for efficient and accelerated AFE in talking portrait systems.
    \item Assessing and discussing the best combinations of talking portrait synthesis systems and AFE systems.

\end{itemize}

The code and resources related to this work are available on GitHub~\cite{whisper-afe-talkingheadsgen}.

\section{Related Work}

The development and research of our avatar is touching upon two main areas: virtual avatars for interviewer training and real-time talking portrait synthesis. 

\subsection{Virtual Avatar for Interview Training}
The evolution of child avatar training systems has advanced investigative interviewing techniques, albeit with varying degrees of automation and efficacy. Early systems, such as those developed by Linnæus University and AvBIT Labs~\cite{dalli2021technological}, primarily utilized prerecorded responses, which constrained interaction dynamics. The LiveSimulation~\cite{roed2023field} enhanced these methods by allowing interaction with a videotaped child, thereby improving open-ended questioning skills~\cite{benson2015evaluation}. Empowering Interviewer Training (EIT)~\cite{pompedda2018training} introduced more dynamic interactions through a rule-based algorithm that facilitated more effective learning with a virtual child. The ViContact~\cite{krause2024prepare} further progressed these methodologies by integrating virtual reality (VR) with automated feedback, thereby enhancing both questioning skills and socio-emotional support.

Building on prior advancements, our previous platform~\cite{salehi2024theoretical} integrated GPT-3 within a Unity framework to simulate child interviews, aiming to improve response dynamism and training effectiveness. Despite these efforts, the system was criticized for the lack of realism in avatar appearance, which detracted from user engagement during interactions~\cite{salehi2024theoretical}. To address this issue, our current work aims to utilize lifelike talking portrait generation in real-world applications, specifically for training in child interview skills, with the expectation that it will improve visual realism.

\subsection{Talking Portrait Synthesis}

In recent years, real-time audio-driven talking portrait synthesis has garnered significant attention due to its applications in digital humans, virtual avatars, and video conferencing. Several approaches have been proposed to balance visual quality, synchronization, and computational efficiency. Live Speech Portrait~\cite{lu2021live} uses auto-regressive predictive coding (APC)~\cite{chung2020generative} for extracting speech information, predicts 3D lip landmarks from audio, and synthesizes video frames through an image-to-image translation network (U-Net). Similarly, RealTalk~\cite{ji2024realtalk} utilizes 3D facial priors and efficient expression rendering modules to achieve precise lip-speech synchronization while preserving facial identity. 

Furthermore, 3D Gaussian Splatting (3DGS)~\cite{kerbl20233d} introduces a point-based rendering technique that uses ellipsoidal, anisotropic Gaussians to represent scenes with high accuracy. GSTalker~\cite{chen2024gstalker} builds on this by incorporating deformable Gaussian splatting, significantly reducing training times and boosting rendering speeds compared to earlier NeRF-based models. Gaussian Talker~\cite{cho2024gaussiantalker} further advances the field by using a Gaussian-based model to generate talking faces with high-quality lip synchronization while reducing computational complexity.

Moreover, NeRF~\cite{mildenhall2021nerf} has recently gained attention for generating talking portraits, given their capacity to capture intricate visual scenes. To enhance system efficiency, RAD-NeRF~\cite{tang2022real} incorporates discrete learnable grids in AD-NeRF~\cite{guo2021ad}, resulting in faster training and inference processes. Building upon this, ER-NeRF~\cite{li2023efficient} utilizes tri-Plane hash representation to minimize hash collisions, leading to faster convergence, while a cross-modal fusion mechanism has been developed to improve lip-speech synchronization. To further refine lip synchronization, GeneFace++~\cite{ye2023geneface++} introduces a dedicated audio-to-motion module within the NeRF-based rendering framework. Additionally, R2-Talker~\cite{ye2023r2} employs a progressive multilayer conditioning approach, improving performance and visual fidelity by integrating hash-grid landmark encoding.

\section{Methodology}
This section introduces the AFE models—Deep-Speech 2, Wav2Vec 2.0, HuBERT, and Whisper—compared in this study and outlines the interactive avatar’s system architecture. 

\subsection{Audio Feature Extraction}

A key aspect of talking-head generation is the ability to capture distinguishing speech features, as this directly affects the synchronization and quality of the audiovisual output. In the proposed framework, feature extraction is conducted using four ASR models: Deep-Speech 2~\cite{amodei2016deep}, Wav2Vec 2.0~\cite{baevski2020wav2vec}, HuBERT~\cite{hsu2021hubert}, and Whisper~\cite{radford2023robust}. Each of these models extracts both acoustic features and language representations from raw audio signals as part of their architecture. The following sections provide further details on these four models.

\subsubsection{Deep-Speech 2}

Deep-Speech 2, developed by Baidu, utilizes bidirectional recurrent neural networks (BRNNs)~\cite{schuster1997bidirectional} alongside convolutional layers to capture context from both past and future frames, enhancing speech recognition accuracy. Key techniques like Batch Normalization and SortaGrad stabilize training, making the model effective across different acoustic conditions, including noisy environments~\cite{amodei2016deep}.

\subsubsection{Wav2Vec 2.0}
Wav2Vec 2.0 is a transformer-based model developed for self-supervised feature extraction directly from raw audio signals~\cite{baevski2020wav2vec}. Initially, the model processes audio waveforms into representations using a convolutional neural network (CNN) paired with a Gaussian Error Linear Unit (GELU) activation function~\cite{hendrycks2016gaussian}, which captures latent speech features across temporal frames $z_1, z_2, \ldots, z_T$. These features are then fed into a transformer network \cite{baevski2019vq}, which is trained through a contrastive loss objective. This loss function enables the model to differentiate between correct and incorrect quantized representations of the audio signal~\cite{jegou2010product}.

This self-supervised training allows Wav2Vec 2.0 to learn rich, contextualized embeddings from unlabeled speech data, effectively building contextual representations across continuous speech and capturing dependencies over the entire sequence of latent representations. This approach reduces the need for hand-engineered features, while leveraging the powerful representations learned by Wav2Vec 2.0, resulting in improved performance for various speech processing applications.

\subsubsection{HuBERT}

Hidden-Unit BERT (HuBERT)~\cite{hsu2021hubert} introduces a self-supervised approach that addresses key challenges in speech processing: handling multiple sound units per utterance, the lack of lexicon during pre-training, and the absence of explicit segmentation of sound units. By applying prediction loss to masked regions, HuBERT learns a combined acoustic and language model from unmasked inputs. Pre-trained on Librispeech~\cite{panayotov2015librispeech} (960 hours) and Libri-light~\cite{kahn2020libri} (60,000 hours), it outperforms previous methods and comes in three sizes: BASE (90M parameters), LARGE (300M), and X-LARGE (1B). HuBERT employs masking similar to SpanBERT~\cite{joshi2020spanbert} and wav2vec 2.0~\cite{baevski2020wav2vec}, using cross-entropy loss on masked and unmasked time steps, encouraging the model to capture both acoustic features and long-range speech patterns.

\subsubsection{Whisper}

In this paper, we propose the use of Whisper Tiny~\cite{radford2023robust}, designed for lightweight applications, offers efficient processing and broad applicability in low-resource or edge environments. With approximately 39 million parameters, it is tailored for real-time applications on less powerful devices while maintaining the core Whisper architecture. This model follows an encoder-decoder Transformer structure, allowing it to perform versatile multilingual transcription, translation, and voice activity detection within a compact and efficient design, making it particularly suitable for talking-head systems and real-time audio-visual synchronization tasks.

Whisper Tiny uses log-Mel spectrograms as input features, derived from 25-millisecond windows with a 10-millisecond stride, which are scaled to a near-zero mean. This spectrogram is processed through Transformer encoder blocks containing convolutional layers with GELU activations~\cite{hendrycks2016gaussian}. These layers capture critical acoustic and linguistic features across languages and environments, leveraging Whisper’s extensive pre-training on multilingual, multitask datasets. This approach provides the model with robust noise resilience and the ability to maintain high accuracy without dataset-specific fine-tuning. 

The Whisper model's design has demonstrated a substantial 80-90\% reduction in processing times compared to alternative models like Deep-Speech, Wav2Vec, and HuBERT, particularly for longer audio clips (see Fig.~\ref{fig:AFE}).

The Whisper model processes raw audio \( A(t) \) by transforming it into a log-Mel spectrogram \( S(f, t) \) as follows:

\[
S(f, t) = \log \left( \sum_{k=0}^{N-1} A(t) \cdot e^{-j2\pi ft} \right)
\]

This spectrogram, capturing core frequency components, is then passed through Whisper's encoder to generate high-dimensional audio embeddings \( E \):

\[
E = \text{WhisperEncoder}(S(f, t))
\]

where \( E \) has shape \( (T_{\text{initial}}, C) \), with \( T \) representing the number of time steps aligned to the visual frames, and \( C = 384 \) as the dimensionality of the feature embedding space. Synchronization with a 25 FPS visual frame rate is achieved by applying a sliding window with parameters \( w = 16 \), stride \( s = 2 \), and padding \( p = 7 \), yielding a final feature matrix with shape \( (750, 16, 384) \). This setup ensures precise temporal alignment across 750 frames over 30 seconds, enhancing the real-time accuracy and fluidity of interactions in talking-head applications.

\subsection{System Architecture}

\begin{figure}[ht!]
    \centering
    \includegraphics[scale=0.87]{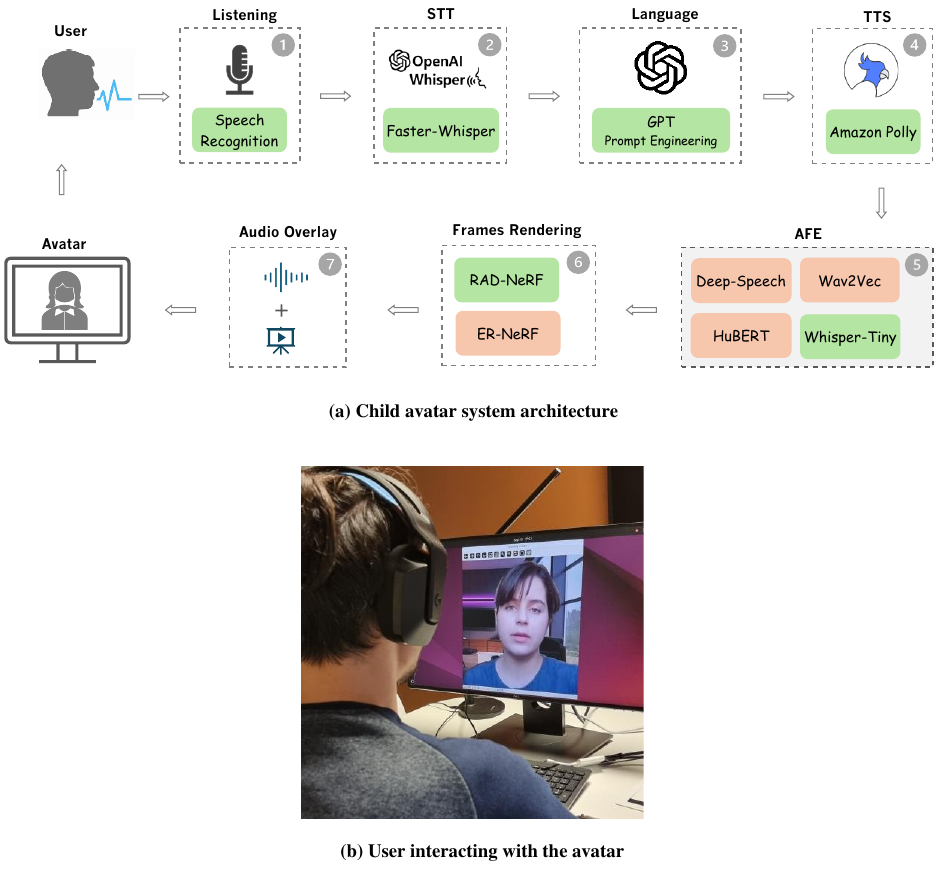}
    
    \caption{\textbf{(a)} System architecture of the interactive child avatar, detailing the integration of key modules: (1) Listening, (2) STT, (3) Language Processing, (4) TTS, (5) AFE, (6) Frames Rendering, and (7) Audio Overlay. This setup simulates natural conversation, allowing the user to interact with the avatar as if communicating with a real person.  
\textbf{(b)} User interaction with the child avatar system.}
    \label{fig:arch}
\end{figure}

The system architecture of our interactive child avatar, as depicted in Figure \ref{fig:arch}, is composed of several modules: Listening, STT, Language, TTS, AFE, Frames Rendering, and Audio Overlay.

The Listening module is the entry point of the system, where user speech is captured through a button-based recording process. Users click to start recording their voice input and click again to stop, saving the recorded audio file. During this recording phase, the avatar remains in a listening state, utilizing a pre-rendered video based on an empty audio file where it exhibits natural, non-verbal behaviors such as blinking and subtle movements while remaining silent. This module employs a speech recognition system to continuously listen for spoken input from the user, initiating the conversion process by passing the audio data to the speech-to-text module.

The STT module utilizes OpenAI's Whisper model~\cite{radford2023robust} to perform real-time conversion of spoken input into text, transcribing the user's speech for further processing. The language module is responsible for generating contextually appropriate and dynamic responses. It leverages GPT for prompt engineering, simulating a child's conversational style. The transcribed text from the STT is processed here, where GPT generates relevant responses tailored to the interaction context. Once the response text is generated, it is passed to the TTS, which uses Amazon Polly \cite{aws_polly} to convert the text into speech. This module maintains the consistency of the avatar's voice with a child's persona.

AFE processes the incoming audio from the TTS module to extract relevant features. To achieve rapid and efficient audio processing that enhances the system's responsiveness, four models were evaluated: Deep-Speech~\cite{amodei2016deep}, Wav2Vec~\cite{baevski2020wav2vec}, HuBERT~\cite{hsu2021hubert}, and Whisper-Tiny~\cite{radford2023robust}.

The Frames Rendering module manages the visual representation of the avatar. It is responsible for rendering frames in real-time based on the processed audio and text input, allowing the avatar to exhibit natural behaviors such as lip-syncing and facial expressions synchronized with the spoken output. We also compared two frameworks, RAD-NeRF~\cite{tang2022real} and ER-NeRF~\cite{li2023efficient}, to identify the most effective solution for this purpose. The final component is the Audio Overlay module, which combines the rendered frames with the audio output.

\section{Experiments}
This section rigorously evaluates various AFE models across two frameworks, focusing on model efficiency, synchronization accuracy, and responsiveness.

\subsection{Experimental Setup}

Datasets, hardware, and configurations are outlined for evaluating AFE model performance in real-time talking portrait synthesis.

\subsubsection{Dataset} 

The dataset used in our experiments comprises a combination of publicly available video datasets and a privately sourced video. We selected three high-definition speech video clips, each with an average duration of approximately 6,700 frames (around 4.5 minutes, as recorded at 25 FPS). The raw videos, originally recorded at their native resolutions (which can also be used in the experiments), were cropped and resized to 512 × 512 pixels. However, the Obama video from AD-NeRF~\cite{guo2021ad}, which was originally processed at 450 × 450 pixels, was used in that resolution without further resizing.

To ensure fairness and reproducibility, two of the video clips used in our experiments were sourced from the Internet. Specifically, we utilized the "Obama" video from the publicly released data of AD-NeRF~\cite{guo2021ad} and the "Shaheen" video. The third video clip, featuring a young girl's speech, was privately sourced to align with the application objectives of the investigative interview. Explicit permission was obtained from the individual for the use of this video in this research.

For each video, the first 91\% frames, along with the corresponding audio, were used as training data, while the final 9\% of the data was reserved for subsequent evaluation, in accordance with previous studies~\cite{guo2021ad, tang2022real, li2023efficient}.

\subsubsection{System Configuration} 

Experiments were conducted on a machine with a 12th Gen Intel(R) Core(TM) i9-12900F CPU, 31 GiB of RAM, and an NVIDIA RTX 4090 GPU with 24 GiB of VRAM, running CUDA 12.4 on an Ubuntu operating system.


\subsection{Real-Time Talking-Head Speed Analysis}
Despite progress in real-time talking-head systems, their widespread use is limited by the complexity of integrating multimodal inputs like audio and facial landmarks. This complexity challenges both synchronization and processing efficiency, which are required for smooth real-time performance.
Given these challenges, we considered it necessary to evaluate the real-time capabilities of existing models to substantiate performance claims. To accurately assess the efficiency of these models, we conducted an experiment using the Obama dataset~\cite{guo2021ad}. The experiment was carried out under identical hardware and CUDA environments across all applicable open-source methods, ensuring a consistent basis for comparison. 

\begin{figure}[ht!]
    \centering
    \includegraphics[scale=0.5]{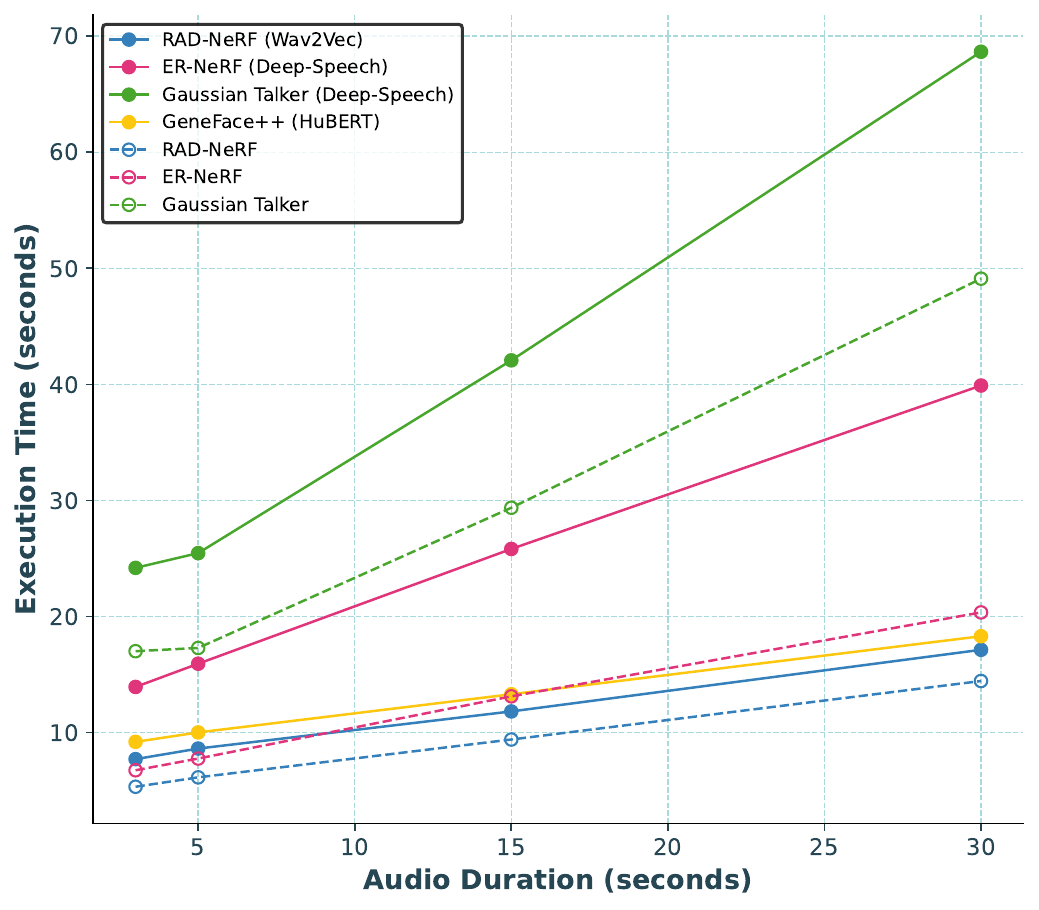}
    \caption{Execution time comparison of open-source real-time talking-head generation models, including RAD-NeRF~\cite{tang2022real}, ER-NeRF~\cite{li2023efficient}, Gaussian Talker~\cite{cho2024gaussiantalker} and GeneFace++~\cite{ye2023geneface++}. The solid lines represent execution times excluding AFE, while the dashed lines indicate execution times that include AFE. }
    \label{fig:models}
\end{figure}

Our findings confirm that a major challenge identified in previous studies is the latency introduced during the AFE process. To better understand the impact of AFE on overall system performance, we measured execution time using two different approaches: one excluding AFE and the other including it. GeneFace++~\cite{ye2023geneface++} is excluded from this comparison, as its audio features are deeply interwoven within the core of the model, making it impossible to measure its performance without AFE. The results, presented in Figure~\ref{fig:models}, illustrate the comparative processing efficiency of these models across different audio durations.

\subsection{AFE Analysis}

Based on the results presented in Figure~\ref{fig:models}, we selected RAD-NeRF~\cite{tang2022real} and ER-NeRF~\cite{li2023efficient} as the frameworks for further experimentation due to their better performance compared to other models. To conduct a systematic comparison of AFE models, we will employ Deep-Speech, Wav2Vec~\cite{baevski2020wav2vec}, HuBERT~\cite{hsu2021hubert}, and Whisper-Tiny~\cite{radford2023robust} within both RAD-NeRF~\cite{tang2022real} and ER-NeRF~\cite{li2023efficient}. The models will be trained from scratch with each AFE configuration to assess their impact on system performance, focusing on factors such as lip synchronization accuracy, visual quality, and overall execution time.
In the following, we divide the analysis into two key aspects—speed and quality—to provide a more detailed evaluation of each AFE configuration's impact on system performance.

\begin{figure}[ht!]
    \centering
    \includegraphics[scale=0.5]{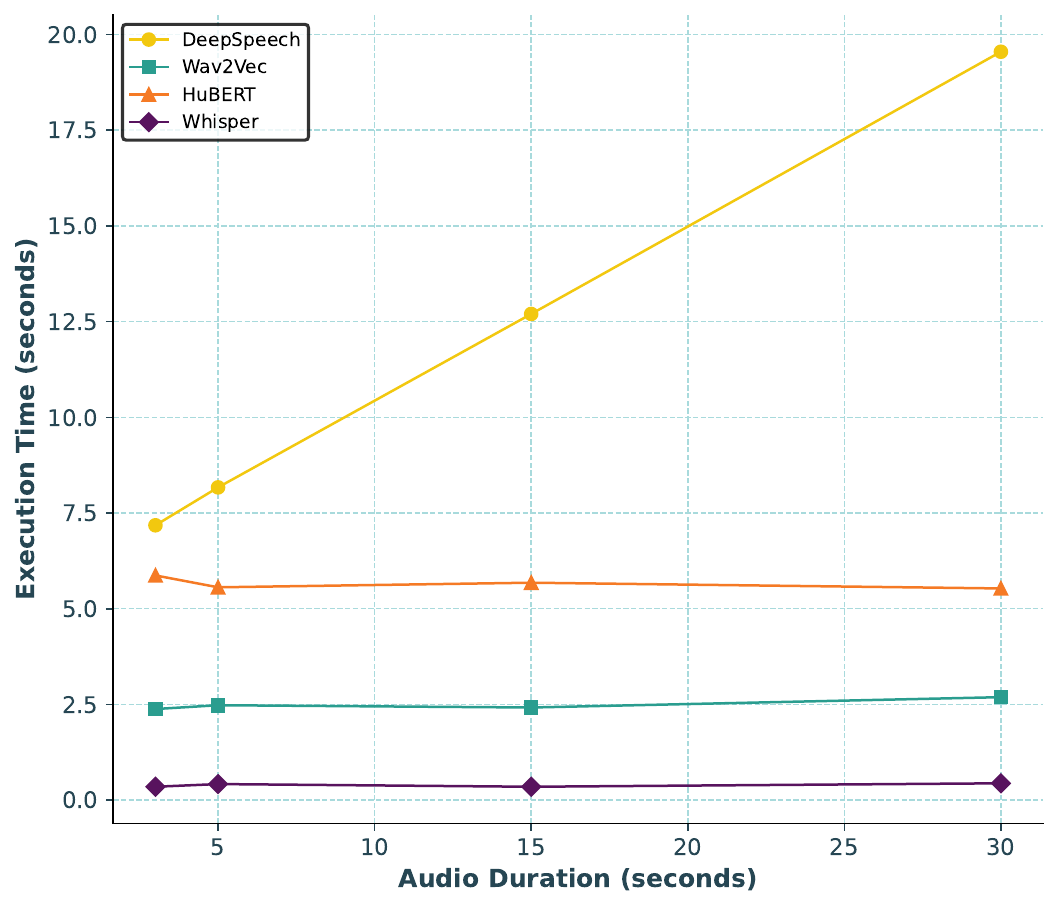}
    \caption{Execution time comparison of different AFE models, including Deep-Speech~\cite{amodei2016deep}, Wav2Vec~\cite{baevski2020wav2vec}, HuBERT~\cite{hsu2021hubert}, and Whisper~\cite{radford2023robust}.}
    \label{fig:AFE}
\end{figure}

\subsubsection{AFE Speed Analysis}

In evaluating the speed of AFEs, we conducted an analysis between Whisper~\cite{radford2023robust} and other well-known AFE models, including Deep-Speech~\cite{amodei2016deep}, Wav2Vec~\cite{baevski2020wav2vec}, and HuBERT~\cite{hsu2021hubert}. The results, illustrated in Figure~\ref{fig:AFE}, reveal that Whisper greatly outperforms the other models across varying audio durations, especially compared to Deep-Speech, which shows increasing execution times as audio duration grows. Whisper consistently achieves notably lower execution times, making it particularly advantageous for conversational systems.

\begin{figure}[ht!]
    \centering
    \includegraphics[scale=0.5]{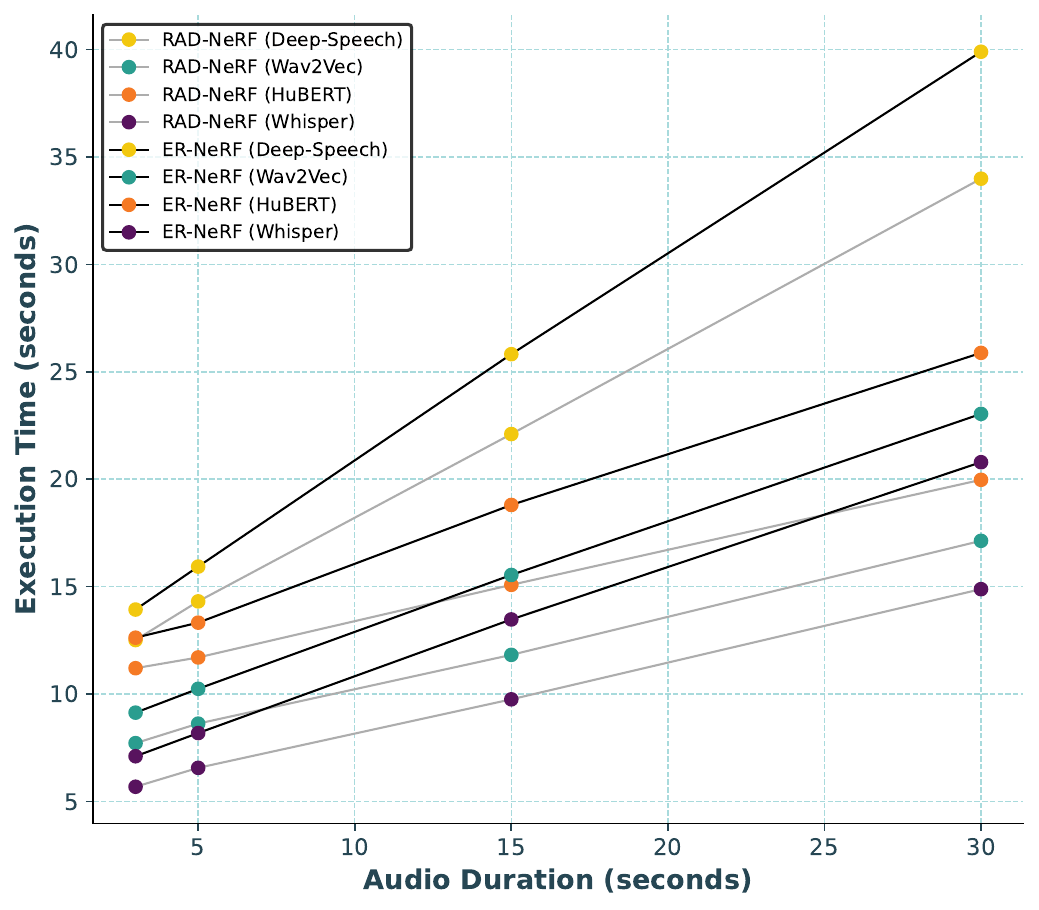}
    \caption{Execution time comparison of RAD-NeRF~\cite{tang2022real} and ER-NeRF~\cite{li2023efficient} across different AFE models.}
    \label{fig:models_AFE}
\end{figure}

Additionally, Figure~\ref{fig:models_AFE} compares the execution times of RAD-NeRF~\cite{tang2022real} and ER-NeRF~\cite{li2023efficient} using various AFE models, including Whisper, Deep-Speech~\cite{amodei2016deep}, Wav2Vec~\cite{baevski2020wav2vec}, and HuBERT~\cite{hsu2021hubert}. The results show the time-saving advantages of Whisper across both models. By integrating Whisper, we achieved a reduction in processing time, which is important for real-time applications such as talking-head generation and interactive avatars. These findings reveal the potential of Whisper as an effective solution for accelerating AFE processes in interactive avatar systems, making them more efficient and responsive.

\subsubsection{AFE quality Analysis}

To further validate utilizing Whisper as AFE, we conducted an evaluation of the system's rendering quality across various settings.
In the self-driven setting, where the ground truth data corresponds to the same identity as the generated output, we employ several quantitative metrics to assess the quality of portrait reconstruction:

\begin{itemize}

    \item \textbf{Peak Signal-to-Noise Ratio (PSNR):} This metric measures the fidelity of the reconstructed image relative to the ground truth. The PSNR is calculated as:
    \[
    \text{PSNR} = 10 \cdot \log_{10} \left( \frac{\text{MAX}_{\text{I}}^2}{\text{MSE}} \right)
    \]
    where $\text{MAX}_{\text{I}}$ is the maximum possible pixel value of the image, and $\text{MSE}$ is the Mean Squared Error between the reconstructed and ground truth images.
    
 \vspace{0.9em}
 
    \item \textbf{Structural Similarity Index Measure (SSIM):} SSIM evaluates structural similarity by considering luminance, contrast, and structure. The formula is:
    \[
    \text{SSIM}(x, y) = \frac{(2\mu_x \mu_y + C_1)(2\sigma_{xy} + C_2)}{(\mu_x^2 + \mu_y^2 + C_1)(\sigma_x^2 + \sigma_y^2 + C_2)}
    \]
    where $\mu_x$ and $\mu_y$ are the average intensities, $\sigma_x$ and $\sigma_y$ are variances, and $\sigma_{xy}$ is the covariance between images $x$ and $y$. $C_1$ and $C_2$ are constants for stability.
    
\vspace{0.9em}

    \item \textbf{Learned Perceptual Image Patch Similarity (LPIPS)~\cite{zhang2018unreasonable}:} Measures perceptual similarity between the generated and ground truth images by calculating the distance between feature representations extracted from a deep neural network. This metric captures differences in visual features that align more closely with human perception than simple pixel-wise comparisons, making it useful for assessing image quality in terms of perceptual fidelity.

\[
\text{LPIPS} = \frac{1}{N} \sum_{i=1}^{N} \| f(x_i^{\text{pred}}) - f(x_i^{\text{truth}}) \|_2
 \]

where \( f(x) \) represents the feature representation of image \( x \) extracted by a neural network (e.g., AlexNet), \( x_i^{\text{pred}} \) and \( x_i^{\text{truth}} \) are the predicted and ground truth images respectively, and \( N \) is the number of patches or feature points compared.

\vspace{0.9em}

    \item \textbf{Landmark Distance (LMD)~\cite{chen2018lip}:} This metric measures the geometric accuracy of facial landmarks by calculating the Euclidean distance between corresponding landmark points in the generated and ground truth images.

\[
\text{LMD} = \frac{1}{N} \sum_{i=1}^{N} \sqrt{(x_i^{\text{pred}} - x_i^{\text{truth}})^2 + (y_i^{\text{pred}} - y_i^{\text{truth}})^2}
 \]

where \( (x_i^{\text{pred}}, y_i^{\text{pred}}) \) and \( (x_i^{\text{truth}}, y_i^{\text{truth}}) \) are the coordinates of the \(i\)-th landmark in the predicted and ground truth images, respectively, and \( N \) is the total number of landmarks.

\vspace{0.9em}

    \item \textbf{Fréchet Inception Distance (FID)~\cite{heusel2017gans}:} FID assesses the similarity between distributions of real and generated images. The formula is:
    \[
    \text{FID} = \mu_r - \mu_g^2 + \text{Tr}(\Sigma_r + \Sigma_g - 2\sqrt{\Sigma_r \Sigma_g})
    \]
    where $\mu$ and $\Sigma$ represent the mean and covariance of features for real (r) and generated (g) images.

\vspace{0.9em}

    \item \textbf{Action Units Error (AUE)~\cite{baltruvsaitis2016openface}:} Measures the accuracy of lower facial muscle movements by calculating the squared differences in action unit intensities between the generated and ground truth images. In this study, we specifically evaluate the lower face region, which is relevant for expressions related to speech and emotion. 

\[
\text{AUE}_{\text{lower}} = \frac{1}{N} \sum_{i=1}^{N} (AU_i^{\text{pred}} - AU_i^{\text{truth}})^2
 \]

where \( AU_i^{\text{pred}} \) and \( AU_i^{\text{truth}} \) are the intensities of the \(i\)-th action unit in the lower face region for the predicted and ground truth data, respectively, and \( N \) is the total number of lower face action units evaluated.

\vspace{0.9em}

    \item \textbf{SyncNet Confidence Score (Sync)~\cite{chung2017out}:} Measures the lip-sync accuracy by evaluating the alignment between audio and lip movements in generated talking-head videos. This metric utilizes SyncNet, which calculates a confidence score based on the similarity between embeddings of audio and video frames. 

\[
\text{Sync} = \frac{1}{N} \sum_{i=1}^{N} \frac{v_i \cdot s_i}{\max(\|v_i\|_2 \cdot \|s_i\|_2, \epsilon)}
 \]

where \( v_i \) and \( s_i \) are the embeddings for the \(i\)-th video and audio frames, respectively, computed by SyncNet. This formula calculates the cosine similarity between the embeddings, giving a score between 0 and 1 for each frame pair. \( N \) is the total number of frame pairs evaluated, and higher average values indicate better lip-sync quality.

\vspace{0.9em}

\end{itemize}

The self-driven evaluation results are shown in Table~\ref{tab:self}, where the utilization of Whisper for AFE leads to a slight improvement in PSNR and LPIPS metrics. Additionally, the FID scores reflect a minor but consistent improvement in the perceptual similarity between generated and real images. However, the most notable improvement is observed in the Sync, which shows better lip synchronization. 

\begin{table*}[htp]
\footnotesize
\centering
\scriptsize
\caption{Quantitative comparison of face reconstruction quality under self-driven synthesis on the same identity's test set.}
\begin{tabular}{cm{1.4cm}m{1 cm}m{0.9cm}m{0.9cm}m{0.9cm}m{0.8cm}m{0.8cm}m{0.8cm}m{1.2cm}}
\toprule
\textbf{Methods} & \textbf{AFE } & \textbf{Dataset} & \textbf{PSNR $\uparrow$} &\textbf{SSIM $\uparrow$}& \textbf{LPIPS $\downarrow$} & \textbf{LMD $\downarrow$} & \textbf{FID $\downarrow$} & \textbf{AUE $\downarrow$} & \textbf{ \textbf{Sync\textsubscript{\textit{conf}}} $\uparrow$}   \\
\midrule

\multirow{12}{*}{\rotatebox{90}{RAD-NeRF \cite{tang2022real}}} & \multirow{4}{*}{Deep-Speech} & Obama & 27.14 & 0.9304 & 0.0738 & 2.675 & 31.29 & 1.995 & 7.171  \\
                                      &                          & Donya & 27.79 & 0.9045 & 0.0917 & 2.750 & 12.82 & 1.911 & 4.720\\
                                      &                         & Shaheen & 30.13 & 0.9314 & 0.0697 & 3.199 & 33.05  & 2.837 &7.330  \\
                                      &                          & \cellcolor{lightgray} \textit{Mean} & \cellcolor{lightgray} 28.35 & \cellcolor{lightgray}0.9221 & \cellcolor{lightgray}0.0784 & \cellcolor{lightgray}2.874 & \cellcolor{lightgray}25.72  & \cellcolor{lightgray} 2.247 & \cellcolor{lightgray}6.407  \\
                                      & \multirow{4}{*}{HuBERT} & Obama & 26.58 & 0.9261 & 0.0769 & 2.762 & 28.78 & 2.006 & 0.563  \\
                                      &                          & Donya & 28.05 & 0.9071 & 0.0868 & 2.518 & 14.25 & 2.511 & 0.365  \\
                                      &                         & Shaheen & 30.45 & 0.9332 & 0.0729 & 3.050 & 35.96 & 3.229 & 0.494 \\
                                      &                          & \cellcolor{lightgray} \textit{Mean} & \cellcolor{lightgray} \textbf{28.36} & \cellcolor{lightgray}0.9221 & \cellcolor{lightgray}0.0788 & \cellcolor{lightgray}\textbf{2.776} & \cellcolor{lightgray}26.33   & \cellcolor{lightgray}  2.582 & \cellcolor{lightgray}0.474  \\
                                      & \multirow{4}{*}{Wav2Vec} & Obama & 26.59 & 0.9268 & 0.0785 & 2.696 & 15.15   & 1.707 & 6.744  \\
                                      &                          & Donya & 27.12 & 0.8972 & 0.0845 & 2.726 & 24.58   & 1.531 & 4.820  \\
                                      &                         & Shaheen & 30.08 & 0.9306 & 0.0698 & 3.221 & 34.77   & 2.966 & 7.946  \\
                                      &                          & \cellcolor{lightgray} \textit{Mean} & \cellcolor{lightgray} 27.93 & \cellcolor{lightgray} 0.9182 & \cellcolor{lightgray}0.0776 & \cellcolor{lightgray}2.881 & \cellcolor{lightgray}24.83   & \cellcolor{lightgray} 2.068 & \cellcolor{lightgray}6.503  \\
                                      & \multirow{4}{*}{Whisper} & Obama & 26.10 & 0.9231 & 0.0723 & 2.573 & 12.67  & 1.693 &7.143  \\
                                      &                          & Donya & 28.65 & 0.9138 & 0.0844 & 2.640 & 26.85 & 1.504 &5.269 \\
                                      &                         & Shaheen & 30.05& 0.9303 & 0.0660 & 3.045 & 29.44  & 2.696 &8.488 \\
                                      &                          & \cellcolor{lightgray} \textit{Mean} & \cellcolor{lightgray} 28.07 & \cellcolor{lightgray}\textbf{0.9224} & \cellcolor{lightgray}\textbf{0.0761} & \cellcolor{lightgray}2.826 & \cellcolor{lightgray}\textbf{24.04} & \cellcolor{lightgray} \textbf{1.964} & \cellcolor{lightgray}\textbf{6.966} \\

\midrule
\multirow{12}{*}{\rotatebox{90}{ER-NeRF \cite{li2023efficient}}} & \multirow{4}{*}{Deep-Speech} & Obama & 26.44 & 0.9339 & 0.0441 & 2.561 & 7.14 &1.923 & 7.201 \\
                                      &                          & Donya & 28.91 & 0.9165 & 0.0605 & 2.647 & 14.59 & 1.874 & 4.722 \\
                                      &                         & Shaheen & 29.92 & 0.9267 & 0.0450 & 2.900 & 16.10 & 2.668 & 8.215 \\
                                      &                          & \cellcolor{lightgray} \textit{Mean} & \cellcolor{lightgray} \textbf{28.16} & \cellcolor{lightgray} \textbf{0.9257} & \cellcolor{lightgray}0.0689 & \cellcolor{lightgray}2.7932 & \cellcolor{lightgray}20.92  & \cellcolor{lightgray} 2.155 & \cellcolor{lightgray}6.712 \\
                                      & \multirow{4}{*}{HuBERT} & Obama & 26.30 & 0.9297 & 0.0473 & 2.758 & 8.33 & 1.711 & 0.300 \\
                                      &                          & Donya & 24.20 & 0.7826 & 0.1255 & 2.545 & 49.81   & 2.284 & 0.408  \\
                                      &                         & Shaheen & 30.45 & 0.9322 & 0.0420 & 2.852 & 16.56 & 3.172& 0.434 \\
                                      &                          & \cellcolor{lightgray} \textit{Mean} & \cellcolor{lightgray} 26.98 & \cellcolor{lightgray} 0.8815 & \cellcolor{lightgray}0.0716 & \cellcolor{lightgray}\textbf{2.718} & \cellcolor{lightgray}24.90   & \cellcolor{lightgray} 2.389 & \cellcolor{lightgray}0.380 \\
                                      & \multirow{4}{*}{Wav2Vec} & Obama & 25.59 & 0.9268 & 0.0497 & 2.645 & 8.83   & 1.704 & 6.616  \\
                                      &                          & Donya & 24.21 & 0.7777 & 0.1509 & 2.754 & 68.20  & 1.730 & 4.403 \\
                                      &                         & Shaheen & 29.81& 0.9245 & 0.0470 & 3.003 & 15.59   & 2.948 &7.917 \\
                                      &                          & \cellcolor{lightgray} \textit{Mean} & \cellcolor{lightgray} 26.53 & \cellcolor{lightgray} 0.8763 & \cellcolor{lightgray}0.0825 & \cellcolor{lightgray}2.800 & \cellcolor{lightgray}30.87  & \cellcolor{lightgray} \textbf{2.127} & \cellcolor{lightgray}6.312  \\
                                      & \multirow{4}{*}{Whisper} & Obama & 26.30& 0.9314 & 0.0462 & 2.501 & 8.06 & 1.797 &7.647 \\
                                      &                          & Donya & 27.36 & 0.9020 & 0.0641 & 2.516 & 14.67 & 1.852 &5.704 \\
                                      &                         & Shaheen & 30.20 & 0.9305 & 0.0434 & 2.935 & 15.61 & 3.030&8.575 \\
                                      &                          & \cellcolor{lightgray} \textit{Mean} & \cellcolor{lightgray} 28.12 & \cellcolor{lightgray} 0.9213 & \cellcolor{lightgray}\textbf{0.0654} & \cellcolor{lightgray}2.7640 & \cellcolor{lightgray}\textbf{19.29} & \cellcolor{lightgray} 2.226 & \cellcolor{lightgray}\textbf{7.308} \\

\bottomrule
\end{tabular}
\label{tab:self}
\end{table*}
\begin{table}[htp]
\footnotesize
\centering
\scriptsize
\caption{Quantitative comparison of audio-lip synchronization under the cross-driven setting.}
\begin{tabular}{ccccc}
\toprule
\textbf{Methods} & \textbf{AFE} & \textbf{Dataset} & \textbf{ \textbf{Sync\textsubscript{\textit{conf}}} $\uparrow$} & \textbf{ \textbf{Sync\textsubscript{\textit{conf}}} $\uparrow$} \\
& & & \textbf{Synthetic} & \textbf{Natural} \\

\midrule

\multirow{12}{*}{\rotatebox{90}{RAD-NeRF \cite{tang2022real}}} & \multirow{4}{*}{Deep-Speech} & Obama & 6.581 & 6.489 \\
                                      &                          & Donya & 4.208 & 3.576  \\
                                      &                         & Shaheen & 6.319 & 5.840  \\
                                      &                          & \cellcolor{lightgray} \textit{Mean} & \cellcolor{lightgray} 5.702 & \cellcolor{lightgray}5.301 \\
                                      & \multirow{4}{*}{HuBERT} & Obama & 5.109 & 0.548 \\
                                      &                          & Donya & 4.750 & 0.680 \\
                                      &                         & Shaheen & 6.670 & 0.741 \\
                                      &                          & \cellcolor{lightgray} \textit{Mean} & \cellcolor{lightgray} 5.509 & \cellcolor{lightgray}0.656 \\
                                      & \multirow{4}{*}{Wav2Vec} & Obama & 6.851 & 7.388 \\
                                      &                          & Donya & 4.593 & 5.017 \\
                                      &                         & Shaheen & 6.837 & 7.303 \\
                                      &                          & \cellcolor{lightgray} \textit{Mean} & \cellcolor{lightgray} \textbf{6.093} & \cellcolor{lightgray}6.569 \\
                                      & \multirow{4}{*}{Whisper} & Obama & 6.884 & 7.137 \\
                                      &                          & Donya & 4.628 & 5.742 \\
                                      &                         & Shaheen & 6.347 & 7.052 \\
                                      &                          & \cellcolor{lightgray} \textit{Mean} & \cellcolor{lightgray} 5.953 & \cellcolor{lightgray}\textbf{6.643} \\

\midrule
\multirow{12}{*}{\rotatebox{90}{ER-NeRF \cite{li2023efficient}}} & \multirow{4}{*}{Deep-Speech} & Obama & 7.306 & 7.224 \\
                                      &                          & Donya & 5.054 & 4.851  \\
                                      &                         & Shaheen & 6.505 & 6.074  \\
                                      &                          & \cellcolor{lightgray} \textit{Mean} & \cellcolor{lightgray} \textbf{6.323} & \cellcolor{lightgray}6.050 \\
                                      & \multirow{4}{*}{HuBERT} & Obama & 5.542 & 0.614 \\
                                      &                          & Donya & 4.822 & 0.480 \\
                                      &                         & Shaheen & 7.068 & 0.581 \\
                                      &                          & \cellcolor{lightgray} \textit{Mean} & \cellcolor{lightgray} 5.810 & \cellcolor{lightgray}0.558 \\
                                      & \multirow{4}{*}{Wav2Vec} & Obama & 7.219 & 7.428 \\
                                      &                          & Donya & 4.115 & 4.155 \\
                                      &                         & Shaheen & 7.145 & 7.242 \\
                                      &                          & \cellcolor{lightgray} \textit{Mean} & \cellcolor{lightgray} 6.159 & \cellcolor{lightgray}6.275 \\
                                      & \multirow{4}{*}{Whisper} & Obama & 7.399 & 8.247 \\
                                      &                          & Donya & 4.651 & 6.486 \\
                                      &                         & Shaheen & 6.148 & 7.093 \\
                                      &                          & \cellcolor{lightgray} \textit{Mean} & \cellcolor{lightgray} 6.066 & \cellcolor{lightgray}\textbf{7.275} \\

\bottomrule
\end{tabular}
\label{tab:cross}
\end{table} 
In the cross-driven setting, the model generates a talking portrait based on audio clips that do not match the identity of the visual data used during training. To create variation, two different audio clips are selected for each dataset, corresponding to the gender and age of the character. One audio clip is synthetic, generated using Amazon Polly \cite{aws_polly}, while the other is a natural recording from a real human, allowing for a fair comparison between synthetic and natural speech.

Because ground truth images corresponding to the same identity are absent, identity-specific metrics are not applicable. Therefore, consistent with prior studies~\cite{tang2022real, li2023efficient}, we use the identity-agnostic Sync score~\cite{chung2017out} as the primary metric to evaluate synchronization between audio and lip movements. This approach allows us to effectively measure the model's performance when direct comparison with ground truth is not feasible.

\begin{figure*}[ht!]
    \centering
    \includegraphics[scale=0.88]{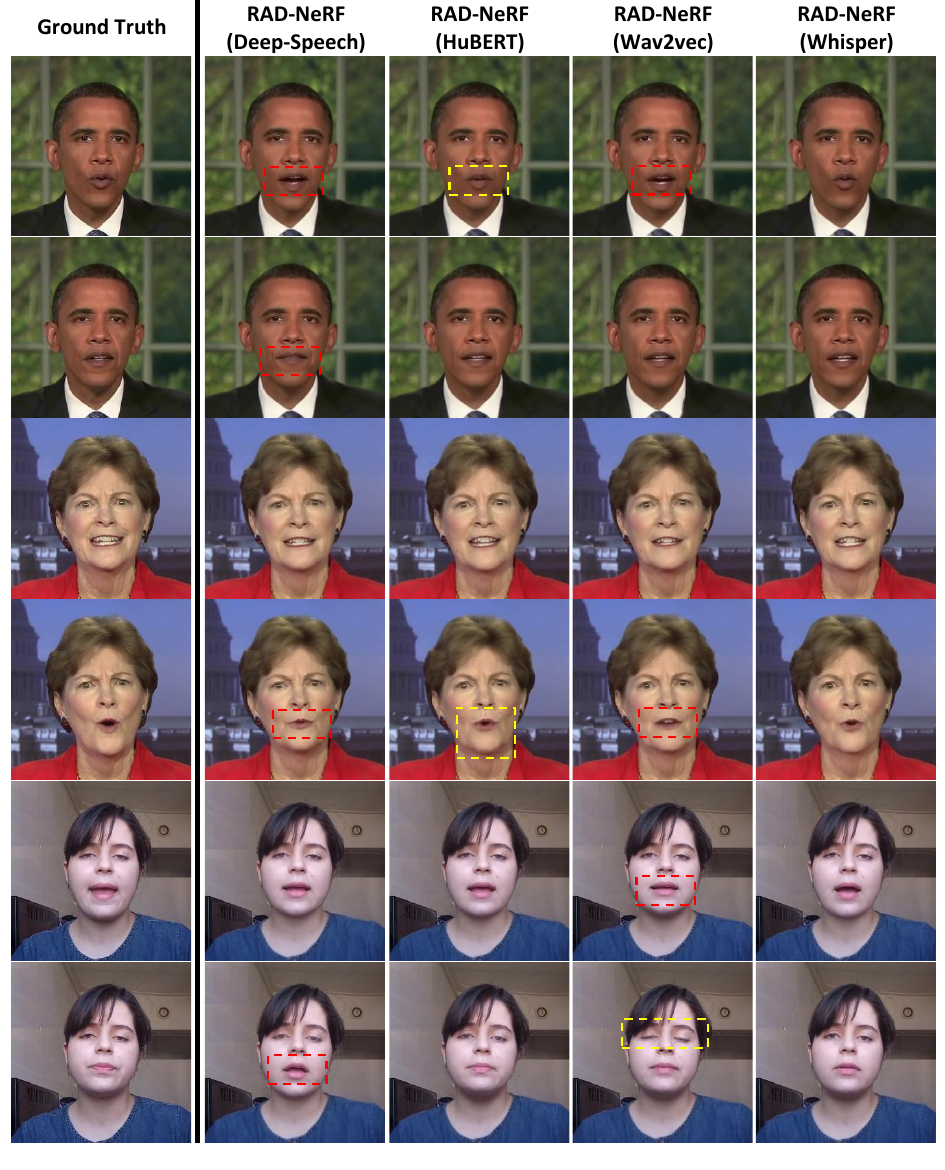}
    \caption{Quality comparison: Examples of visualizations of RAD-NeRF~\cite{tang2022real} under the self-driven setting, based on two frames extracted from each video illustrating typical challenges. Yellow boxes highlight areas of noisy image quality, while red boxes indicate regions with inaccurate lip synchronization.}
    \label{fig:f_rad}
\end{figure*}

\begin{figure*}[ht!]
    \centering
    \includegraphics[scale=0.88]{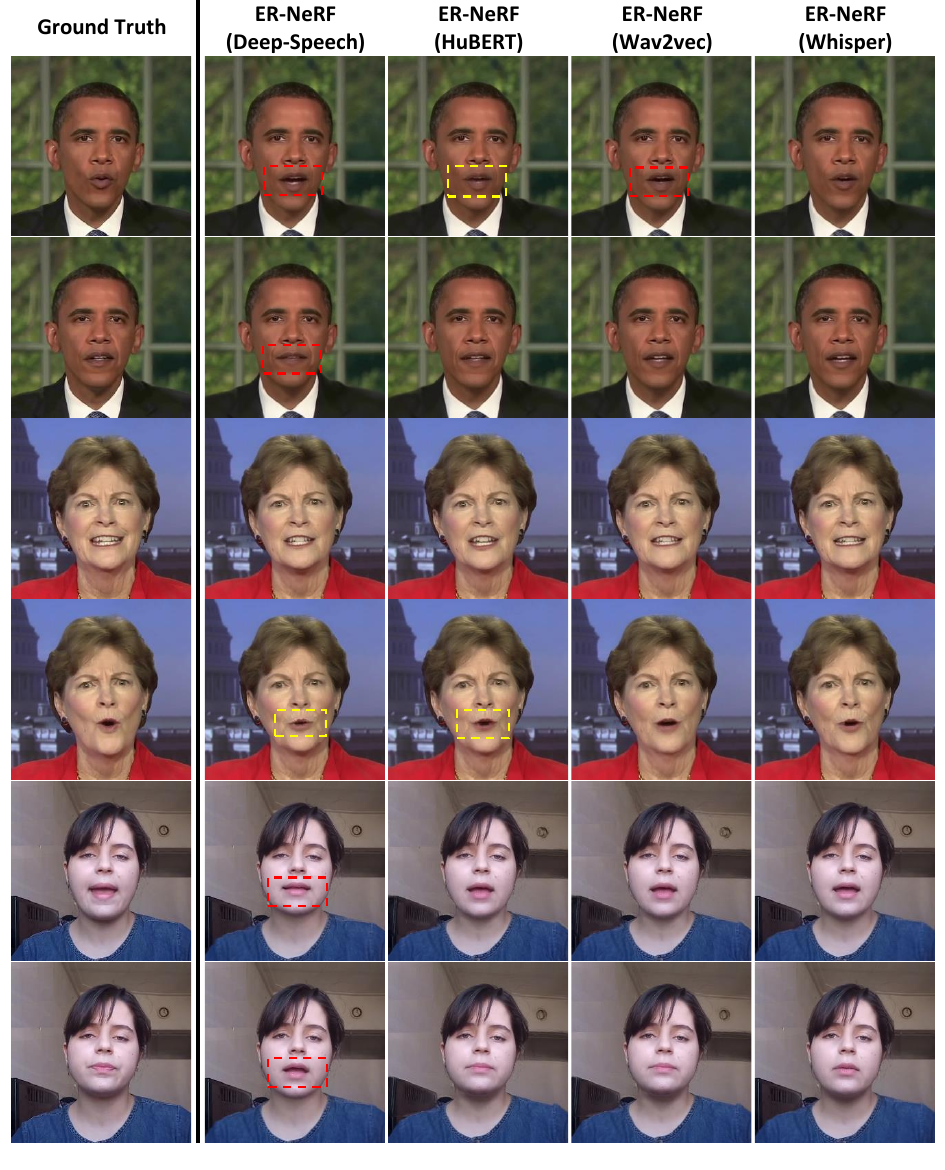}
    \caption{Quality comparison: Examples of visualizations of ER-NeRF~\cite{li2023efficient} methods under the self-driven setting, based on two frames extracted from each video illustrating typical challenges. Yellow boxes highlight areas of noisy image quality, while red boxes indicate regions with inaccurate lip synchronization.}
    \label{fig:f_er}
\end{figure*}

The cross-driven results, as shown in Table~\ref{tab:cross}, indicate that models utilizing Whisper generally outperform others, particularly in terms of SyncNet scores. However, Whisper's superior performance is not consistent across all scenarios. While it shows particular strength in handling human voices, its effectiveness diminishes when dealing with bot-generated voices. Specifically, in our experiments, the bot's voice was generated at a slower pace and interspersed with more pauses. This slower speech setting may have unintentionally benefited other models, which made them perform better relative to Whisper in this context.

Our experimental results reveal that HuBERT consistently underperforms in lip synchronization, with the lowest Sync scores among all the models tested. Given these results, particularly the poor performance in cross-driven settings (e.g., Sync scores as low as 0.564), HuBERT appears inadequate for applications requiring precise lip-sync. Future work should consider incorporating subjective evaluations to further assess the perceptual quality and user satisfaction, which may provide a more thorough understanding of its limitations.

Furthermore, we compared the subjective quality and lip-sync performance of two models, RAD-NeRF~\cite{tang2022real} and ER-NeRF~\cite{li2023efficient}, shown in Figure~\ref{fig:f_rad} and Figure~\ref{fig:f_er}, respectively, using four distinct AFEs under self-driven settings. Each model’s output was evaluated against ground truth to assess its accuracy in audio-driven 3D reconstruction and lip-sync fidelity. For a more comprehensive view of the results, please refer to the supplemental video available on our GitHub repository~\cite{whisper-afe-talkingheadsgen}.

ER-NeRF and RAD-NeRF demonstrated varying degrees of fidelity across the AFEs, particularly in achieving accurate lip-sync with the audio input. Generally, Whisper achieved the closest approximation to ground truth across both ER-NeRF and RAD-NeRF, exhibiting strong performance in synchronizing lip movements with audio, especially during pronounced lip articulations, such as the wide-open mouth movements observed when pronouncing sounds like "wa." This precise alignment enhances the realism of the output, particularly in sequences requiring dynamic mouth shapes.

Wav2vec and HuBERT provided reasonable approximations but showed slight misalignments in lip-sync. Deep-Speech, while effective, displayed the greatest variance from ground truth, with notable lip-sync discrepancies, indicating less robust performance in these NeRF-based reconstructions.

\subsection{System Responsiveness Analysis}

Table~\ref{tab:tab_system} outlines execution times per component, with the Listening component excluded due to its dependence on user input duration. The analysis is segmented into raw execution time and the percentage contribution of each stage to the overall system latency, providing a granular view of where time is consumed in the process of generating the interactive avatar’s responses.

\begin{table*}[htp!]
\centering
\scriptsize
\caption{Execution Time Analysis of Each Component in the System Architecture as Depicted in Figure \ref{fig:arch}. \textbf{AA Tokens:} Number of tokens in the avatar’s answer. \textbf{AA Duration (sec)}: Length of the avatar’s answer in seconds.}
\begin{tabular}{>{\centering\arraybackslash}m{0.8cm} >{\centering\arraybackslash}m{1.5cm} | m{1.7cm} >{\centering\arraybackslash}m{0.6cm} >{\centering\arraybackslash}m{1cm} >{\centering\arraybackslash}m{0.6cm} >{\centering\arraybackslash}m{0.6cm} >{\centering\arraybackslash}m{1.2cm} >{\centering\arraybackslash}m{0.95cm} >{\centering\arraybackslash}m{0.6cm}}
\toprule
       \textbf{AA Tokens} & \textbf{AA Duration (sec)}  &  & \textbf{STT} & \textbf{Language} & \textbf{TTS} & \textbf{AFE} & \textbf{Frame Rendering} & \textbf{Audio Overlay} &  \textbf{SUM} \\
\toprule
\multirow{2}{*}{1} &  \multirow{2}{*}{0.41} & \textbf{Exe. Time (sec)} & 0.06 & 0.80 & 0.22 & 0.29 & 4.05 & 0.14 & 5.56 \\
                     &     &    \textbf{\% of Total} & 1.08\% & 14.39\% & 3.96\% & 5.22\% & 72.84\% & 2.52\% & 100\% \\

\midrule
 \multirow{2}{*}{8} &  \multirow{2}{*}{1.69} & \textbf{Exe. Time (sec)} & 0.07 & 0.96 & 0.33 & 0.28 & 4.75 & 0.16 & 6.55 \\
                            &   & \textbf{\% of Total} & 1.07\% & 14.66\% & 5.04\% & 4.27\% & 72.52\% & 2.44\% & 100\% \\

\midrule
 \multirow{2}{*}{14} & \multirow{2}{*}{3.63} & \textbf{Exe. Time (sec)}  & 0.07 & 2.45 & 0.44 & 0.28 & 5.45 & 0.14 & 8.83 \\
                               &    & \textbf{\% of Total} & 0.79\% & 27.74\% & 4.98\% & 3.17\% & 61.71\% & 1.59\% & 100\% \\

\midrule
 \multirow{2}{*}{21} & \multirow{2}{*}{5.08} & \textbf{Exe. Time (sec)}  & 0.1 & 2.27 & 0.44 & 0.28 & 5.88 & 0.17 & 9.14 \\
                           &    & \textbf{\% of Total} & 1.09\% & 24.83\% & 4.81\% & 3.06\% & 64.36\% & 1.86\% & 100\% \\      

\midrule
 \multirow{2}{*}{30} & \multirow{2}{*}{6.55} & \textbf{Exe. Time (sec)}  & 0.06 & 2.76 & 0.49 & 0.27 & 6.58 & 0.15 & 10.31 \\
                           &    & \textbf{\% of Total} & 0.58\% & 26.77\% & 4.75\% & 2.62\% & 63.84\% & 1.45\% & 100\% \\

\midrule
 \multirow{2}{*}{39} & \multirow{2}{*}{9.55} & \textbf{Exe. Time (sec)}  & 0.09 & 1.95 & 0.78 & 0.28 & 7.52 & 0.18 & 10.8 \\
                            &    & \textbf{\% of Total} & 0.83\% & 18.06\% & 7.22\% & 2.59\% & 69.63\% & 1.67\% & 100\%  \\

\midrule
\multirow{2}{*}{50} & \multirow{2}{*}{12.16} & \textbf{Exe. Time (sec)}  & 0.07 & 2.08 & 0.55 & 0.28 & 8.65 & 0.18 & 11.81 \\
                          &    & \textbf{\% of Total}  & 0.59\% & 17.61\% & 4.66\% & 2.37\% & 73.24\% & 1.52\% & 100\% \\

\bottomrule
\end{tabular}
\label{tab:tab_system}
\end{table*}

To evaluate the system's responsiveness, we tested various lengths of avatar answers. This approach allowed us to assess how the system handles different interaction complexities, ranging from brief exchanges like \textit{"Hi"} and \textit{"I'm OK"} to more extended dialogues, such as \textit{"Um... at Jenny's house, we play a lot. It's nice there... but, well, I don't really like talking about the pool. It wasn't very fun last time"}. The results indicate that while the overall system performs efficiently, the frame rendering stages are identified as the most time-consuming, due to the computational demands of generating high-quality, real-time visual output that matches the synchronized audio input.

Conversely, the integration of the Whisper model as the AFE component has proven to be much faster than the conventional AFE models. Whisper’s improved processing speed has substantially reduced the time required to extract and synchronize audio features, contributing to an overall faster system performance.

\section{Discussion and Future Work}

The results support the value of Whisper~\cite{radford2023robust} as a robust and efficient AFE in real-time talking portrait synthesis, especially for applications that require responsiveness. In comparing execution times across AFE models, including Deep-Speech~\cite{amodei2016deep}, Wav2Vec~\cite{baevski2020wav2vec}, HuBERT~\cite{hsu2021hubert}, and Whisper~\cite{radford2023robust}, it is evident that Whisper consistently achieves lower execution times. This reduction in processing latency is particularly important for real-time applications where delays can disrupt user engagement~\cite{salehi2024theoretical}. The faster response enabled by Whisper allows smoother interactions in applications like interactive avatars, which improves both the user experience and system reliability.

The performance differences among the AFE models are significant, particularly regarding speed. Whisper’s streamlined architecture, which efficiently processes raw audio into high-dimensional feature embeddings, contrasts with more resource-intensive models like Deep-Speech, which tends to slow down with longer audio durations. This efficiency translates directly to reduced system latency, which is important for maintaining the immediacy expected in real-time avatar interactions. While speed is a significant advantage, Whisper also stands out in synchronization accuracy and visual quality, albeit with smaller differences. Higher SyncNet confidence scores for Whisper indicate superior lip-sync accuracy, a key factor for creating lifelike avatars. This precision contributes to system realism, helping to mitigate the uncanny valley effect~\cite{mori2012uncanny} that can disrupt user immersion in avatar interactions~\cite{salehi2022more}.

Whisper excels in speed and is optimized for multilingual, multitask scenarios, efficiently handling diverse audio inputs with minimal latency. Its feature encoding integrates seamlessly with visual frame rendering, ensuring rapid visual adjustments in real-time speech applications. This design provides an edge over other models by maintaining low execution times across audio durations, which is particularly valuable for dynamic avatars requiring consistent, timely responses.

These findings suggest that Whisper can improve the responsiveness and realism of interactive avatars. However, despite Whisper’s efficiency, the overall system still experiences latency due to other computationally intensive components, such as frame rendering. Future work could focus on optimizing these components or utilizing advanced solutions like NVIDIA's Avatar Cloud Engine (ACE)\footnote{https://developer.nvidia.com/ace}, a suite of technologies that combines generative AI and hardware acceleration specifically for creating realistic digital humans. By leveraging tools like ACE, which use powerful NVIDIA GPUs to accelerate processes like speech recognition, natural language processing, and real-time facial animation, system responsiveness can be further improved. Additionally, conducting broader user studies with professionals in the field could provide a deeper understanding of the system’s impact on practical outcomes, such as user training and engagement. Expanding this research to include real-world evaluations with subjective user feedback will offer valuable insights into perceived realism, usability, and overall user satisfaction. Incorporating user-centric metrics will provide a more comprehensive view of the system’s effectiveness in high-engagement applications, such as investigative interview training. Through these advancements, AI-driven avatars may achieve greater utility and realism, enhancing their role as effective tools in training applications.

\section{Conclusion}

This study addressed the latency challenges associated with AFE, which has hindered the practical deployment of real-time talking portrait systems in real-world applications. By integrating the Whisper model—a high-performance ASR system—into our framework, we achieved notable reductions in processing delays. These optimizations not only increased the overall responsiveness of the interactive avatars but also improved the accuracy of lip-syncing, making them more applicable for immersive training applications.

Our findings affirm Whisper’s capability to meet real-time demands, particularly in applications requiring responsive interactions and minimal delay. This efficiency is important for training environments such as CPS, where timely and realistic interactions can greatly impact training efficacy. By achieving these improvements, Whisper-integrated systems emerge as promising solutions for a variety of real-time applications, including virtual assistants, remote education, and digital customer service platforms.

\end{document}